\documentclass[a4paper,aps,superscriptaddress,floatfix,nofootinbib,twocolumn,longbibliography]{revtex4-1}

\usepackage{amsmath, amsthm, amssymb} 
\usepackage{algpseudocode} 
\usepackage{comment} 
\usepackage{graphicx}
\usepackage{dcolumn}
\usepackage{bm}
\usepackage{hyperref}
\usepackage[mathlines]{lineno}
\usepackage{float}

\theoremstyle{definition}

\begin{document}
\title{
Influence Maximization: Divide and Conquer
}

\author{Siddharth Patwardhan}
\affiliation{Center for Complex Networks and Systems Research, Luddy School of Informatics, Computing, and Engineering, Indiana University, Bloomington, Indiana 47408, USA}

\author{Filippo Radicchi}
\affiliation{Center for Complex Networks and Systems Research, Luddy School
  of Informatics, Computing, and Engineering, Indiana University, Bloomington,
  Indiana 47408, USA}
\email{filiradi@indiana.edu}

\author{Santo Fortunato}
\affiliation{Indiana University Network Science Institute (IUNI)}
\affiliation{Center for Complex Networks and Systems Research, Luddy School of Informatics, Computing, and Engineering, Indiana University, Bloomington, Indiana 47408, USA}
\email{santo@indiana.edu}

\begin{abstract}

The problem of influence maximization, i.e., finding the set of nodes having maximal influence on a network, is of great importance for several applications. In the past two decades, many heuristic metrics to spot influencers have been proposed. Here, we introduce a framework to boost the performance of any such metric. 
The framework consists in dividing the network into sectors of influence, and then selecting the most influential nodes within these sectors.
We explore three different methodologies to find sectors in a network: graph partitioning, graph hyperbolic embedding, and community structure. 
The framework is validated with a systematic analysis of real and synthetic networks.
We show that the gain in performance generated by dividing a network into sectors before selecting the influential spreaders increases as the modularity and heterogeneity of the network increase. Also, we show that
the division of the network into sectors can be efficiently performed in a time that scales linearly with the network size, thus making the framework applicable to large-scale influence maximization problems.

\end{abstract}

\maketitle


\section{Introduction}

The spread of news, ideas, rumours, opinions, and awareness in social networks is generally analyzed in terms of processes of information diffusion ~\cite{notarmuzi2022universality, newman2018networks,banerjee2020survey,granell2013dynamical}. 
A well-established feature of this type of processes on real, heterogeneous networks is that a small fraction of nodes may have a disproportionately large influence over the rest of the system~\cite{morone2015influence,erkol2019systematic,banerjee2020survey}. Therefore, influence maximization (IM)-- the problem of finding the optimal set of nodes that have the most influence or the largest collective reach on the network-- is central for potentially many applications~\cite{domingos2001mining, banerjee2020survey}. 

Kempe \textit{et al.} were the first to formalize the IM problem~\cite{kempe2003maximizing}. They showed that the problem is NP-hard, and that solutions to the IM problem can only be approximated. Also, they proposed a greedy optimization algorithm guaranteeing a solution that is within a factor $(1 - 1/e) \simeq 0.63$ from the optimal solution for two main classes of spreading models.  Greedy optimization consists in building the set of influential spreaders in a network sequentially by adding one spreader at a time to the  set. At each stage of the algorithm, the best spreader is chosen as the node, among those outside the current set of optimal spreaders, that generates the largest increment in the influence of the set of spreaders.  Importantly, the gain in influence that a candidate spreader could bring is estimated by adding it to the current set of already selected spreaders, and simulating numerically the spreading process. This procedure, although computationally expensive, allows for properly assessing the combined influence that multiple spreaders usually have in a network. The original recipe by Kempe \textit{et al.} can be applied to relatively small networks only. Followup studies further improved upon the complexity of the greedy algorithm proposed by Kempe \textit{et al.} allowing for the study of IM problems in larger settings~\cite{chen2009efficient,chen2014cim,banerjee2020survey,shang2017cofim,bozorgi2016incim,bagheri2016efficient}. 
Speedup is also possible by first dividing the network into sectors, and then performing greedy optimization within each sector separately~\cite{chen2014cim,shang2017cofim,bozorgi2016incim,banerjee2020survey,bagheri2016efficient}. In these approaches, sectors are generally identified in terms of network communities. Finding communities in networks is a task that can be performed in a time that grows linearly with the network size~\cite{fortunato10}.
However, since 
these algorithms still rely on the estimation of the influence function via numerical simulations,
they can only be used to deal with IM problems on networks of moderate size.

As more efficient alternatives, several purely topological metrics of node centrality were proposed to quantify the influence of the nodes~\cite{chen2009efficient,morone2015influence,newman2014eigen,brin1998apagerank,lu2016h}. The assumption behind this approach is that a topological centrality metric is a good proxy for dynamical influence. 
As the computation of a network centrality metric does not involve simulating the actual spreading process, centrality-based algorithms can be applied to study the IM problem in large-scale networks. However, their performance in approximating solutions to IM problem is systematically worse than that of the greedy algorithm~\cite{erkol2019systematic}. 

A common drawback of centrality-based algorithms is assuming that each seed acts as an independent spreader in the network so that the influence of a set of spreaders is given by the sum of the influence of each individual seed. This is clearly a weak assumption. For example, it is well known that even in the case of simple contagion models like the independent cascade model, the best strategy is not choosing highly influential nodes in the same closely connected neighborhood~\cite{schoenebeck2022think}, but choosing sufficiently distant nodes~\cite{holme2017inf}. Two main ways of alleviating this issue are considered in the literature. A first way consists in defining an adaptive version of the centrality metric at hand, so that the effect of the already selected spreaders is discounted from the estimation of the influence of the nodes under observation. This trick is able to greatly improve the performance of even basic degree centrality, whose adaptive version excels in performance~\cite{erkol2019systematic}. A second way proposed by Chen {\it et al}. is first partitioning the network into sectors, and then estimating nodes' influence within their own sectors~\cite{chen2020node2vec}.  The rationale behind this procedure is that sectors represent relatively independent parts of a network, thus selecting seeds from different sectors represents a straightforward way of reducing the overlap between portions of the network that multiple spreaders are able to influence. The rationale is similar to the one used in greedy optimization performed on network communities~\cite{chen2014cim,shang2017cofim,bozorgi2016incim,banerjee2020survey,bagheri2016efficient}, however, sectors in Chen \textit{et al.} are obtained by  clustering nodes on the basis of their node2vec embedding~\cite{grover2016node2vec}. One of the advantages of using geometric embedding instead of community structure is the possibility of having full control on the number of sectors used in the division of the network. On the other hand, identifying sectors in an high-dimensional space as the one generated by node2vec is computational expensive. Further in the procedure by Chen \textit{et al.}, the number of sectors is set equal to the number of spreaders that should be identified, requiring therefore to find sectors afresh whenever the size of the seed set is varied. The result is an algorithm that does not scale well with 
the system size.

In this paper, we generalize and combine the above ideas into a scalable approach. We propose a pipeline consisting in dividing the network into sectors and then choosing influential spreaders based on the division of the network into sectors. Scalability is obtained by imposing the number of sectors to be independent from the number of spreaders. We explore three different methodologies to divide the network into sectors, namely graph partitioning, graph hyperbolic embedding, and community structure. The first two methods allow us to identify sectors in the graph in a time that grows linearly with the network size. The use of centrality metrics like adaptive degree centrality that also can be computed in linear time allows us to produce solutions to the IM problem in large networks. Hyperbolic embedding requires instead a time that grows quadratically with the network size, but allows for a flexible and straightforward way of identifying network sectors. The method can be used only in sufficiently small networks. 

We systematically validate our approach on a large corpus of real-world networks, demonstrating its effectiveness in approximating solutions to the IM problem. Furthermore, we leverage the Lancichinetti-Fortunato-
Radicchi (LFR) network model~\cite{lancichinetti2008benchmark} to show that the method is particularly useful in solving IM problems on modular and  heterogeneous networks.

\begin{figure}[!htb]
    \includegraphics[width=\linewidth]{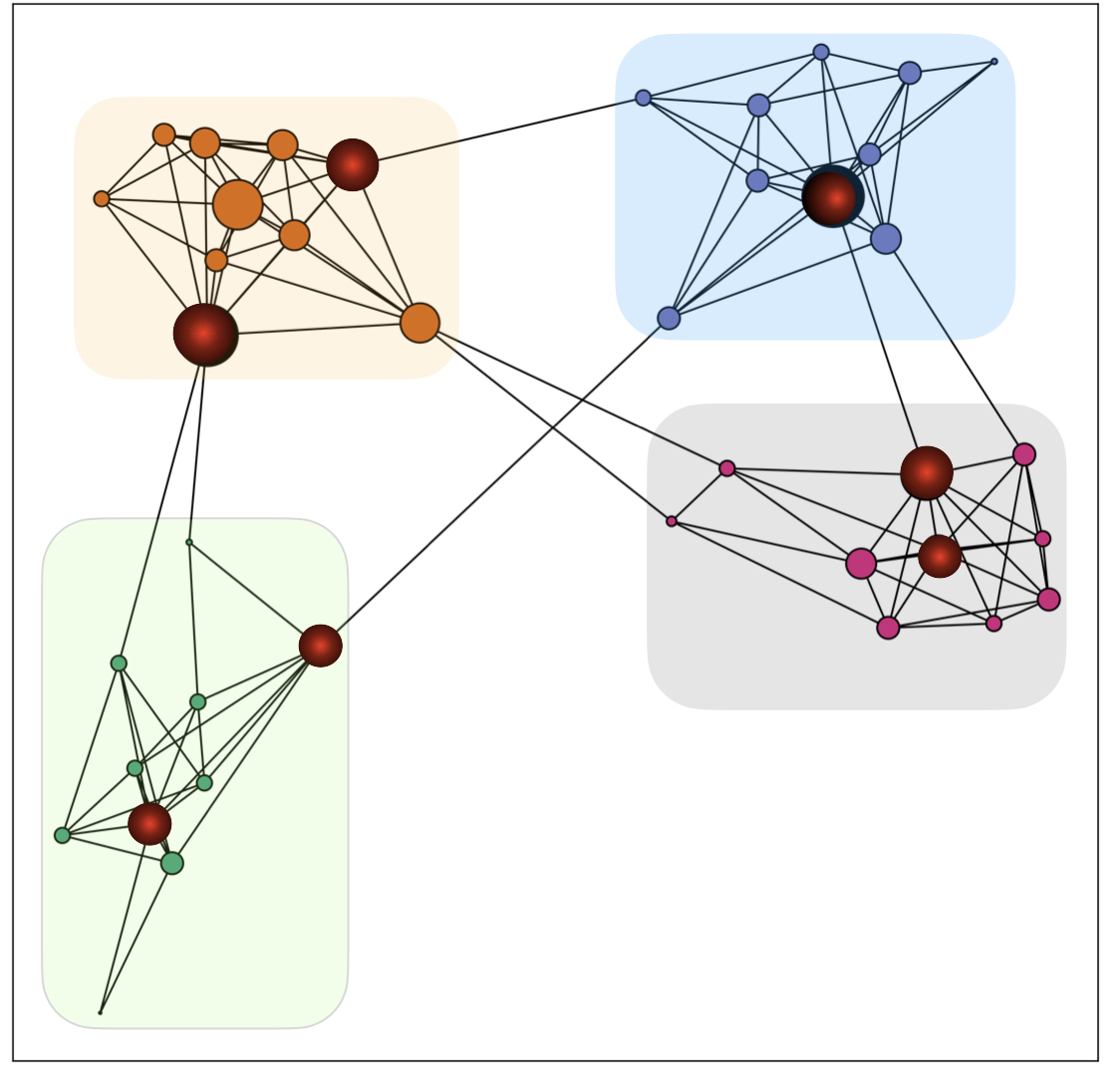}
    \label{fig:1}
    \caption{\textbf{The divide-and-conquer approach to influence maximization.} 
    The network is first divided into sectors of influence, here represented by different colors. Each influential spreader is chosen by first randomly picking a sector, and then selecting a node within the sector, that is not yet part of the set of spreaders, according to some criterion, typically the value of a centrality score. The operation is iterated until a desired number of spreaders is selected. The size of nodes in the figure is proportional to their degrees, here used to proxy nodes' influence. Seven influential spreaders, depicted as bold circles, are selected from the four available sectors.
    }
\end{figure}

\section{Methods}
\subsection{Networks}

\subsubsection{Real networks}
We take advantage of a corpus of $52$ undirected and unweighted real-world networks. Sizes of these networks range from $N=500$ to $N=26,498$ nodes.
The upper bound on the maximum size of the networks analyzed is due to the high complexity of the greedy optimization algorithm, which we use as the baseline for estimating the performance of the other algorithms.
We consider networks from different domains. Specifically, our corpus of networks include social, technological, information, biological, and transportation networks. Details about the analyzed networks can be found in the Appendix.

\subsubsection{LFR model}

To systematically analyze the dependence of the proposed algorithm's performance on the modularity and the heterogeneity of the network structure, we use the LFR network model~\cite{lancichinetti2008benchmark}, commonly adopted as benchmark for community detection algorithms~\cite{fortunato2016community}. The LFR model allows us to generate synthetic networks with power-law distributions of degree and community size. Parameters of the model are the power-law exponent of the degree distribution $\tau_1$, the average degree $\langle k \rangle$, the maximum degree $k_{max}$, the power-law exponent of the community size distribution $\tau_2$, and the mixing parameter $\mu$, which is the average fraction of neighbors outside the community of a node. Low values of $\mu$ indicate well separated and pronounced communities; the larger $\mu$ the less strong the community structure is.

\subsection{Independent cascade model}

In this work, we focus our attention on the Independent Cascade Model (ICM)  which is one of the 
most studied spreading models in the context of influence maximization (IM)~\cite{kempe2003maximizing}. The ICM is a discrete-time contagion model, similar in spirit to the Susceptible-Infected-Recovered model~\cite{pastor2015epidemic}. 
In the initial configuration, all nodes are in the susceptible state, except for the nodes in the set of spreaders that are in the infected state. At a given time step, each infected node first attempts to infect its susceptible neighbors with probability $p$, and then recovers. 
Recovered nodes do no longer participate in the dynamics. The dynamics proceeds by repeating the previously described iteration over the newly infected nodes.
The spreading process stops once there are no infected nodes remaining in the network. The influence of the set of spreaders is quantified as the size of the outbreak, i.e., the number of nodes that are found in the recovered state at the end of the dynamics. Clearly, this number may differ from realization to realization of the model due to the stochastic nature of the spreading events. The IM problem consists in finding the set of spreaders leading to the largest average value of the outbreak size~\cite{kempe2003maximizing}. The optimization is constrained by the number of nodes that can compose the set of spreaders. The typical setting in practical applications consists in finding a small set of spreaders in a very large network.  

As a function of the spreading probability $p$, the ICM displays a transition from a non-endemic regime, where the size of the outbreak is small compared to the network size, to an endemic regime, where the outbreak involves a large portion of the nodes in the network. The IM problem is particularly challenging and interesting around the point where such a change of regime occurs. We define it as the pseudo-critical value $p^*$ of the ordinary bond-percolation model on the network. Specifically, $p^*$ represents the threshold between the non-endemic and endemic regimes for the ICM started from one randomly chosen seed; this fact follows from the exact mapping of critical SIR-like spreading to bond percolation on networks~\cite{grassberger1983critical}. We stress that each network is characterized by a different $p^*$ value; the numerical estimation of a network's $p^*$ is performed using the Newman-Ziff algorithm \cite{newman2000efficient,radicchi2015predicting}.

\subsection{The divide-and-conquer algorithm}
The input of our algorithm is an unweighted and undirected network $G = (V,E)$, with set of nodes $V$ and set of edges $E$. We denote the size of the network as $N = |V|$. The algorithm requires also to choose the number $k$ of desired influential spreaders, and the number $S$ of sectors used to divide the network. The divide-and-conquer (DC) algorithm consists of two main components (see Figure~\ref{fig:1}). First, we divide the network into $S$ sectors, or vertex subsets, $V_1, V_2,..., V_S$. We have $V = \bigcup_{i=1}^S V_i$ and $V_i \cap V_j = \emptyset$ for all $i \neq j$. Second, we form the set of $k$ influential spreaders by adding one node at a time to the set. Starting from an empty set, at each of the $k$ iterations, we first select a random sector, and then pick the most influential node in the sector that is not already included in the set of spreaders. One can use any suitable methodology to divide the network and any suitable centrality metric to select influential spreaders from the sectors. Clearly, for $S=1$ no actual division of the network into sectors is performed. In this case, the selection of influential spreaders is made relying on the centrality metric scores only, thus according to the standard procedure used in the literature~\cite{erkol2019systematic}. For $S=N$, seed nodes are randomly selected.

We note that the above procedure is conceptually identical to the one introduced by Chen {\it et al.}~\cite{chen2020node2vec}. However, there are a few important practical differences. First, Chen {\it et al.} consider high-dimensional node2vec embeddings only~\cite{grover2016node2vec}. node2vec requires a non-trivial calibration of several hyperparameters that is known to be essential for task performance, but adds significant computational burden to the procedure~\cite{zhang2021systematic}. Also, the high-dimensionality of the node2vec embedding space makes the identification procedure of the sectors non trivial. Finally, Chen {\it et al.} impose $S = k$, with one seed selected per sector. This fact implies that increasing the seed set from $k$ to $k+1$ requires redefining the sectors afresh, an operation that requires a time that grows at least linearly with the network size $N$. Since in IM problems one typically uses a number of spreaders proportional to the size of the system~\cite{erkol2019systematic}, the resulting complexity of the algorithm is at least quadratic.

\subsubsection{Dividing the network}

We consider three possible methods of dividing a network into sectors: (i) graph partitioning, (ii) graph hyperbolic embedding, and (iii) community structure. Below, we briefly summarize each of these methods.


{\bf Graph partitioning}  
consists in splitting a graph into an arbitrarily chosen number of sectors
of roughly equal size, such that the total number of edges lying between the corresponding subgraphs is minimized \cite{bichot2013partitioning, karypis1997metis}.
To perform graph partitioning, we take advantage of  METIS~\cite{karypis1997metis}, i.e., the algorithm that implements the multilevel partitioning technique introduced in Refs.~\cite{karypis1998multilevel} and~\cite{karypis1998multilevelk}. 
The computational time of METIS grows as $S\, N$~\cite{karypis1997metis}.


{\bf Graph hyperbolic embedding} is another representation that allows to divide a network into sectors. Here, sectors are given by groups of close-by nodes in vector space. The geometric representation  in hyperbolic space offers full control on the size and number of sectors that can be formed. Such a division can be performed efficiently relying on the angular coordinates of the nodes only. This fact greatly simplifies the identification of sectors compared to higher-dimensional embeddings such as those considered by Chen {\it et al.}~\cite{chen2020node2vec}. We take advantage of the algorithm named Mercator to map nodes into the hyperbolic disk~\cite{garcia2019mercator}.
Mercator does not have hyperparameters, so no calibration is needed.
On the weak side, Mercator performs the embedding of a network with $N$ nodes in a time proportional to $N^2$, clearly limiting the application of the method to small/medium-sized networks.


{\bf Community structure} also can be leveraged to divide the network into sectors by assuming that communities represent sectors. This idea is clearly inspired by the IM algorithms of Refs.~\cite{chen2014cim,shang2017cofim,bozorgi2016incim,banerjee2020survey,bagheri2016efficient}. Roughly speaking, the community structure of a network is a partition of the graph into groups of nodes having higher probability of being connected to each other than to members of other groups~\cite{fortunato2016community}. Plenty of algorithms are available on the market to find community structure in networks. Here, we take advantage of the Louvain algorithm~\cite{blondel2008fast}. 
Louvain is known for its speed (i.e., computational complexity grows linearly with the number of nodes in the network). It has major limitations~\cite{fortunato2016community}, but our procedure does not demand high accuracy in the detection of communities and we do not expect results to be dramatically different if one used another community detection algorithm. Compared to graph partitioning and graph embedding,  
an apparent issue
in using community structure to define sectors of influence is that community detection algorithms do not generally offer the possibility to control for the size and the number of communities. 

\subsubsection{Conquering the network}

We proxy the influence of nodes using topological centrality metrics.
This procedure is similar to the one used by Chen {\it et al.}~\cite{chen2009efficient}, but different from the one considered in Refs.~\cite{chen2014cim,shang2017cofim,bozorgi2016incim,banerjee2020survey,bagheri2016efficient}. We limit our attention only to metrics that can be computed in a time that grows almost linearly with the network size. We rely on the following metrics.

{\bf Adaptive degree centrality} is a simple, but powerful metric for approximating nodes' influence in IM problems~\cite{chen2009efficient, erkol2019systematic}. The metric is designed for the sequential construction of a set of spreaders; in such a procedure, the adaptive degree centrality of a node is given by the total number of connections that a node has towards other nodes that are not included in the current set of spreaders. Unless otherwise specified, all our implementations of the DC algorithm rely on adaptive degree centrality. 

{\bf Collective influence} is a natural generalization of adaptive degree centrality~\cite{morone2015influence}. When computed for node $i$, the metric is a function of the degrees of the nodes that are at shortest-path distance $\ell$ from node $i$. $\ell$ is a free integer parameter. For $\ell = 0$, the metric reduces to adaptive degree centrality. We report results obtained for $\ell =2$, which is a standard setting in IM problems~\cite{erkol2019systematic}. 

{\bf Eigenvector centrality} measures a node’s importance while considering the relative importance of its neighbors. It assigns relative scores to all nodes in the network such that an edge to a more central node contributes more to a node's score than an edge to a less central node \cite{bonacich1972factoring}.

\subsection{Notation}

For sake of compactness, we adopt the following notation for the various methods used to  approximate solutions of the IM on networks. The strategy used to proxy the influence of individual nodes is denoted by lower-case letters. Specifically, we use $\textrm{g}$ to denote greedy optimization, and $\textrm{r}$ to indicate random selection. For the metrics of centrality we use $\textrm{a}$ to indicate adaptive degree centrality,  $\textrm{c}$ for collective influence, and $\textrm{e}$ for eigenvector degree centrality. If the above metrics of centrality are used within our proposed DC scheme, then we use a notation where the lower-case letter of the centrality metric is preceded by an upper-case letter indicating the specific method used to define sectors. We use $\textrm{P}$ to denote graph partitioning, $\textrm{E}$ for hyperbolic graph embedding, and $\textrm{C}$ for community structure. For example, the method $\textrm{m}$ that leverages hyperbolic graph embedding to boost the performance of adaptive degree centrality is denoted as $\textrm{m} = \textrm{Ea}$; the method $\textrm{m}$ that uses community structure in combination with eigenvector centrality is denoted as $\textrm{m} = \textrm{Ce}$.

\subsection{Metrics of performance}

We measure the performance of each method using a metric similar to the one defined in Ref.~\cite{erkol2019systematic}. Indicate with $\mathcal{X}^{(k)}_\textrm{m} = \{x_\textrm{m}^{(1)}, x_\textrm{m}^{(2)}, \ldots, x_\textrm{m}^{(k)}\}$  the set of the $k$ seeds identified by the method. We estimate the average value of the outbreak size generated by the set $\mathcal{X}^{(k)}_m$ by performing $500$ simulations of the ICM. Indicate this quantity as $O^{(k)}_\textrm{m}$. We then compute the sum
\begin{equation}
A_\textrm{m} = \sum_{k=1}^{11} \,  O^{(r_k)}_\textrm{m} \; ,
\label{eq:auc}
\end{equation}
where $r_k = \lfloor [0.01 + (k-1) 0.004]  \, N \rfloor$ and $\lfloor \cdot \rfloor$ is the floor function. 
This metric approximates  the overall performance of the method $\textrm{m}$ in building sets of influential spreaders of sizes ranging from $1\%$ to $5\%$ of the network size. The increment $0.004$ only serves to divide this range in $10$ bins of equal size.
We finally compute the ratio
\begin{equation}
R_\textrm{m} = \frac{A_\textrm{m}}{A_\textrm{g}} \; .
\label{eq:norm_auc}
\end{equation}
According to the above metric, the performance of the method is measured relatively to the baseline provided by greedy optimization, i.e., $A_\textrm{g}$. The normalization serves to make values of the metric comparable across networks of different size.

\begin{figure}[!htb]
    \includegraphics[width=\linewidth]{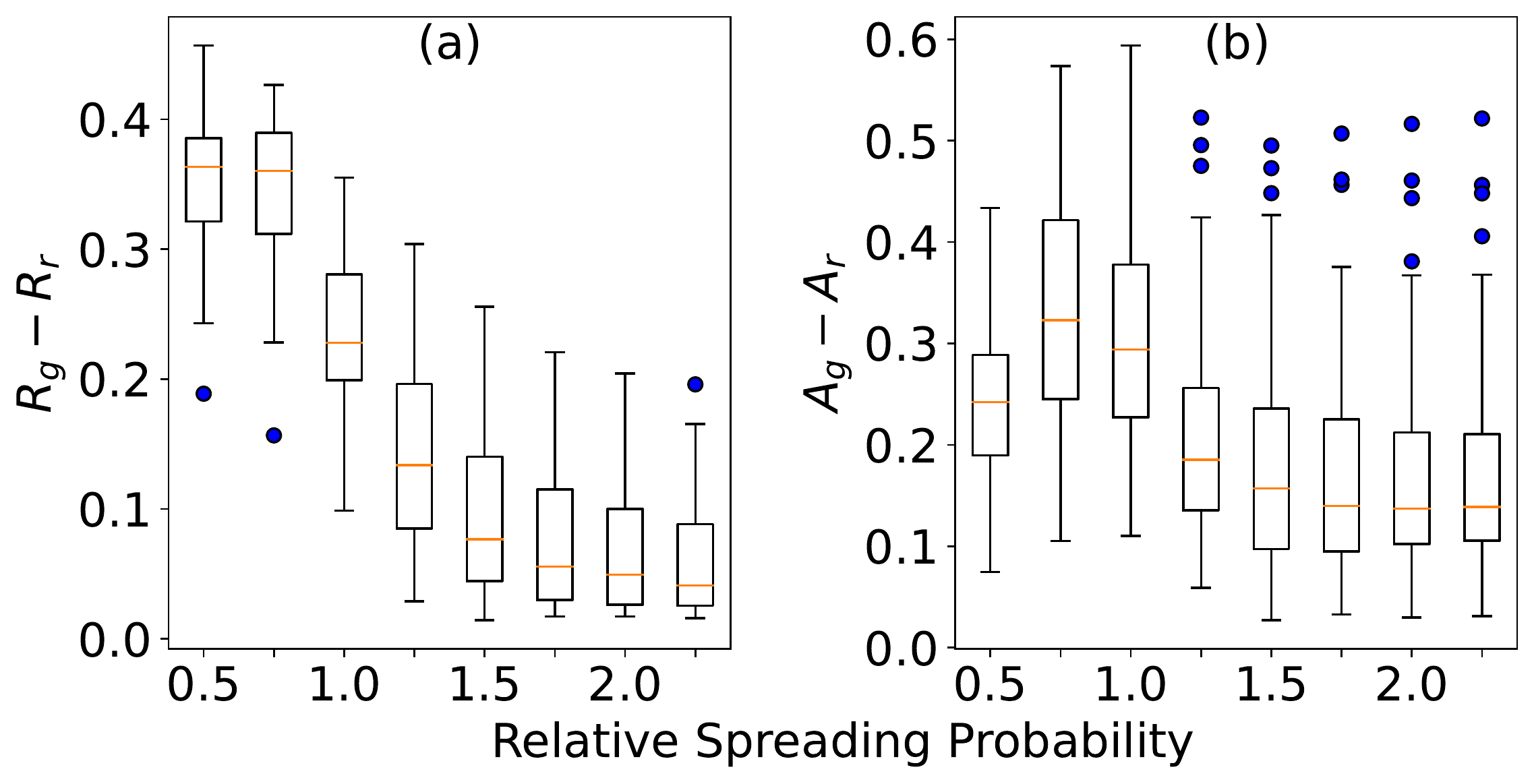}
    \caption{{\bf Influence maximization on real-world networks.} (a) For each real network, we evaluate the critical spreading probability $p^*$. Set of spreaders are identified either using greedy optimization or random selection. We then evaluate the performance metric of 
    Eqs.~(\ref{eq:norm_auc}) using $500$ ICM realizations for each value of the spreading probability $p$. We plot the difference $R_\textrm{g} - R_\textrm{r}$ as a function of the relative spreading rate, i.e., the ratio $p/p^*$. Results stem from the $52$ real networks considered in our analysis. The orange line in the boxplot represents the median value. The boxes show the first and third quartiles of the data, and the whiskers extend from the box to include the 1.5 inter-quartile range. The blue points are the data points not included within the error bars. 
    (b) Same as in panel (a), but we plot $A_\textrm{g} - A_\textrm{r}$, as defined in Eq. (\ref{eq:auc}), as a function of the ratio $p/p^*$.}
    \label{fig:2}
\end{figure}

\section{Results}

\subsection{Spreading probability}

The value of the spreading probability $p$ has a considerable impact on the outcome of the spreading process, and consequently on the properties of the associated IM problem. Trivially, for $p=0$ or $p=1$, any strategy for choosing the set of spreaders is equivalent in terms of performance. The problem becomes non trivial in the vicinity of the pseudo-critical point $p^*$, where uncertainty in the outcome of the spreading process is maximal if seeds are chosen at random, but appropriately setting the initial condition of the spreading should strongly determine the actual size of the outbreak. In this section, we emphasize the importance of studying the spreading process near the critical threshold $p^*$. 
We show results for the $52$ networks in our corpus in Figure \ref{fig:2}. 
We plot $R_\textrm{g} - R_\textrm{r}$
as a function of the relative spreading probability, i.e., $p/p^*$.
Note that each network has its own $p^*$ value.
The curve $R_\textrm{g} - R_\textrm{r}$ assumes high values for $p \leq p^*$ and drops quickly for $p \geq p^*$. The discrepancy between the random and greedy selection strategies is also well characterized by the difference  $A_\textrm{g} - A_\textrm{r}$, which peaks around $p \simeq p^*$. Assuming that a generic algorithm for IM displays a performance that is bounded above by the greedy algorithm and bounded below by random selection, we deduce that $p \simeq p^*$ is the regime of the dynamics where different algorithms to approximate the IM problem should be compared.

\begin{figure}[!htb]
    \centering
    \includegraphics[width=\linewidth]{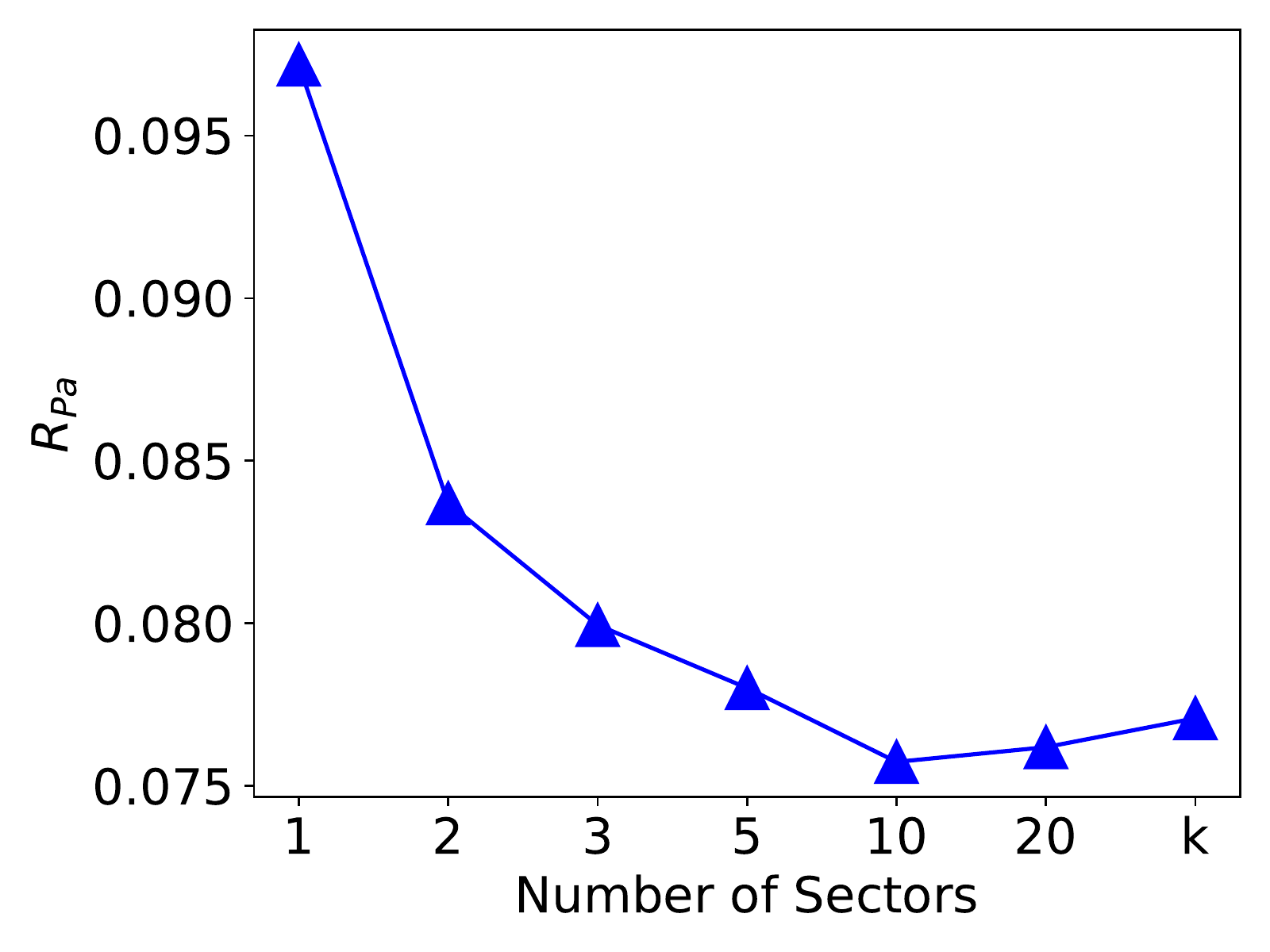}
    \caption{\textbf{Sectors of influence in real-world networks.}
    We display the average performance of the DC approach based on graph partitioning and adaptive degree centrality, i.e., $R_\textrm{Pa}$ [Eq.~(\ref{eq:norm_auc})], as a function of the number of sectors $S$. $S=k$ indicates that sectors are varied between $\lfloor 0.01 N \rfloor$ to $\lfloor 0.05 N \rfloor$ as we compute the metric of  Eq.~(\ref{eq:auc}). Performance values shown in the figure are averaged over the 52 networks in our corpus. The outbreaks sizes were obtained from $500$ independent simulations of the ICM.  
    }
    \label{fig:3}
\end{figure}

\begin{figure*}[!htb]
    \centering
    \includegraphics[width=\linewidth]{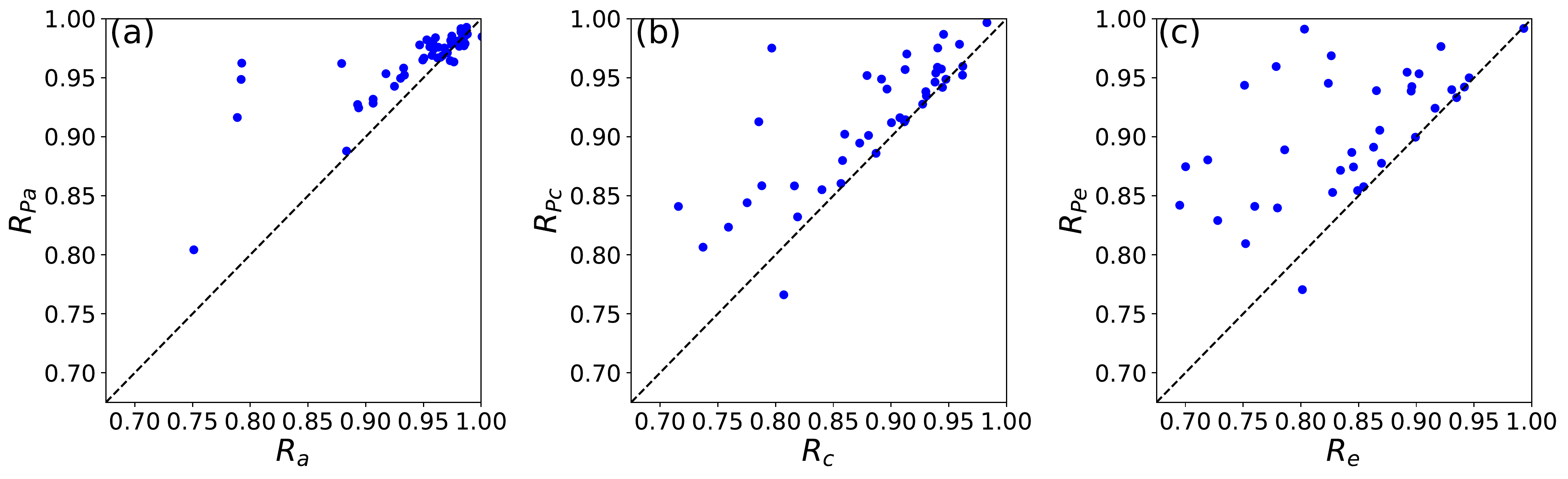}
    \caption{{\bf Performance of the divide-and-conquer algorithm on real networks.} (a) Each point in the graph is a real-world network. Their coordinates are given by the estimated ratios $R_\textrm{Pa}$ and $R_\textrm{a}$, representing the performance of the divide-and-conquer algorithm leveraging adaptive degree using ten sectors and the one using only one sector, respectively. The dashed line indicates equal performance of the two methods. (b) Same as in panel (a), but for $R_\textrm{Pc}$ and $R_\textrm{c}$, i.e., influence of nodes is estimated using collective influence (parameter $\ell=2$ in this tests). (c) Same as in panel (a), but but for $R_\textrm{Pe}$ and $R_\textrm{e}$, i.e.,  influence of nodes is estimated using eigenvector centrality.}
    \label{fig:4}
\end{figure*}

\begin{figure*}[!htb]
    \centering
    \includegraphics[width=\linewidth]{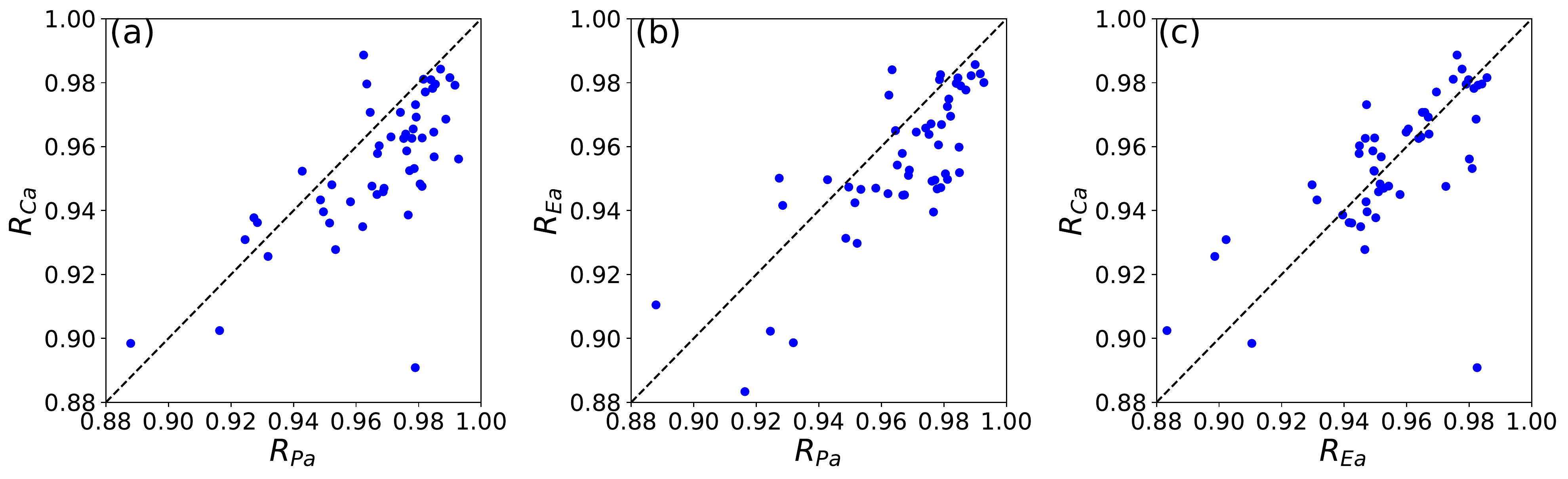}
    \caption{
    {\bf Performance of the divide-and-conquer algorithm on real networks.} (a) Each point in the graph is a real-world network. Their coordinates are given by the estimated $R_\textrm{Pa}$ and $R_\textrm{Ca}$ values, representing the performance of the divide-and-conquer (DC) algorithm leveraging graph partition and community structure, respectively. In both cases, after the network is divided into sectors, the influcenc of individual nodes is estimated using adaptive degree centrality. The dashed line indicate equal performance between the two methods. (b) Same as in panel (a), but comparing $R_\textrm{Pa}$ and $R_\textrm{Ea}$, i.e., the performance the DC algorithm based on graph hyperbolic embedding. (c) Same as in (a) and (b), but comparing $R_\textrm{Ca}$ and $R_\textrm{Ea}$.
    }
    \label{fig:5}
\end{figure*}

\subsection{Number of sectors}

The proposed DC approach involves first dividing the nodes into $S$ subsets,
and then determining the most central nodes within the various sectors. The choice of the parameter $S$ influences the performance and the efficiency of the approach.

We note that the conquer component of the algorithm has computational complexity that is independent of $S$. For example, computing adaptive degree centrality requires a time that grows as $N \log N$~\cite{morone2016collective}. 
However, computing other centrality metrics may be more demanding than that.

The computational complexity of the divide component of the algorithm depends on the specific method utilized. Finding communities with Louvain requires a time that grows slightly super-linearly with the network size $N$~\cite{blondel2008fast}; the number of communities $S$ is not a freely tunable parameter, thus the computational time does not have any explicit dependence on it. Embedding a graph in hyperbolic space with Mercator requires a time that grows quadratically with the system size~\cite{garcia2019mercator}. Once the embedding is given, the $S$ sectors can be found by first sorting the angular coordinates of the nodes, thus requiring a time that grows slightly super-linearly with $N$, and then obtaining $S$ slices in a time that grows linearly with $S$.
 The computational complexity of METIS grows as $S\, N$~\cite{karypis1997metis}; it is therefore advisable choosing $S$ growing at most logarithmically with the network size $N$ in order to avoid significant computational burden. 
 
 We find that using a value of $S$ between $10$ and $20$ yields the optimal relative outbreak size for the real networks in our corpus. Moreover, in many networks, we see that any value of $S>1$ gives us some advantage over $S=1$. In this paper we set the value of $S=10$, unless specified otherwise. We justify this choice of $S$ by comparing the metric $R_\textrm{Pa}$ defined in Eq. (\ref{eq:norm_auc}) for different values of $S$. We compare the performance for $S=1,2,3,5,10,20$ in Figure \ref{fig:3}. In the figure, we include also results obtained by setting $S$ equal to the number of $k$ influencers. Please note that this number is not constant, but varied between $\lfloor 0.01 N \rfloor$ to $\lfloor 0.05 N \rfloor$ while estimating Eq.~(\ref{eq:auc}). We see that $S=10$ is the best choice for our approach. 

\begin{figure*}[!hbt]
    \includegraphics[width=0.9\linewidth]{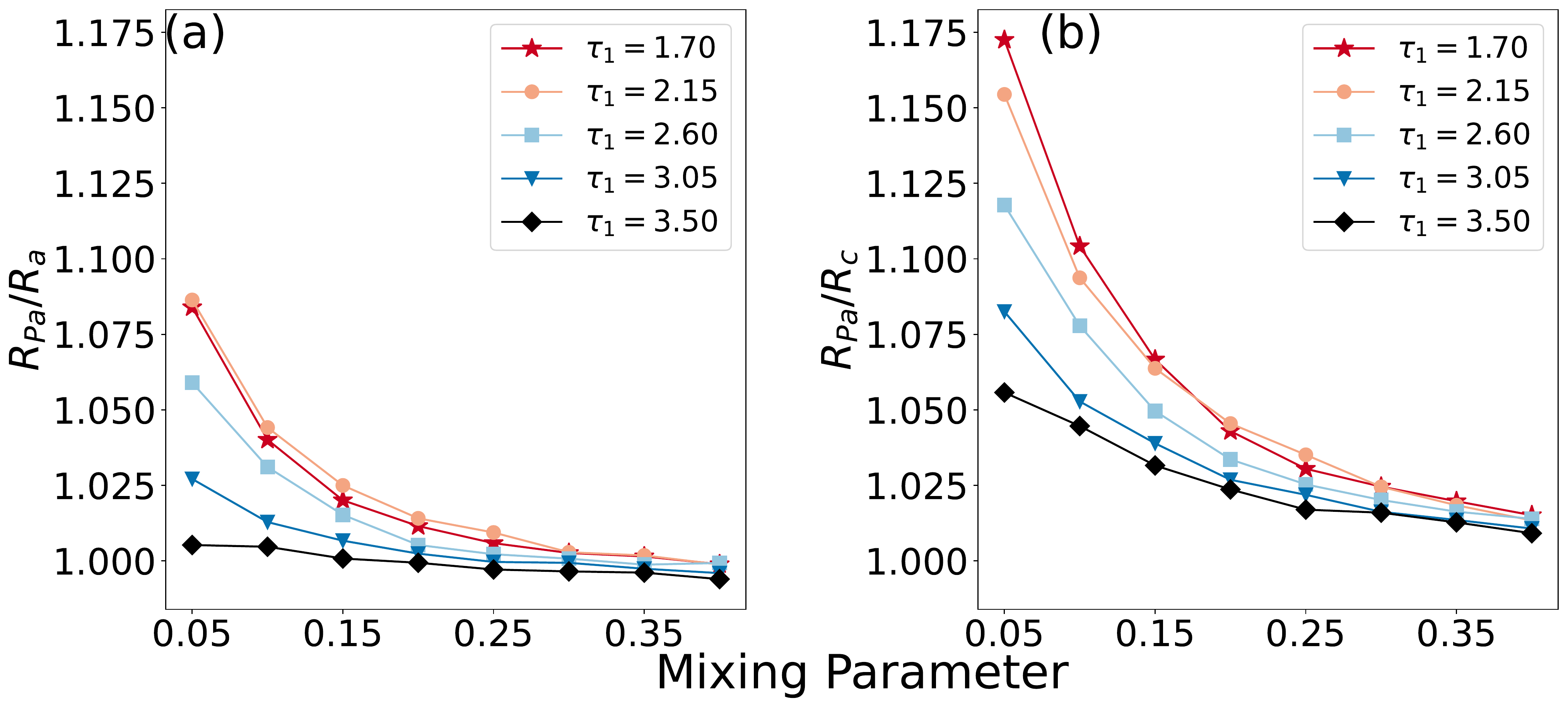}
    \caption{\textbf{Performance of the divide-and-conquer algorithm on synthetic networks.}
    (a) We generate synthetic networks using the LFR model~\cite{lancichinetti2008benchmark}. We consider networks with $N=1,000$ nodes, community size power-law exponent $\tau_2 = 1$, average degree $\langle k \rangle = 10$, and maximum degree $k_{max} = 70$.  We plot the ratio $R_\textrm{Pa}/R_\textrm{a}$ as a function of the mixing parameter $\mu$. Different curves correspond to different values of the degree exponent $\tau_1$. (b) We consider the same networks as in panel (a), but we plot $R_\textrm{Pa}/R_\textrm{c}$ as a function of $\mu$.
    }
    \label{fig:6}
\end{figure*}


\subsection{Real-world networks}

We consider critical ICM dynamics, and monitor how the size of the outbreak changes as a function of the size of the seed set. We use different variants of the DC algorithm based on graph partitioning, where the influence of individual nodes is estimated based on adaptive degree centrality, collective influence and eigenvector centrality, respectively. We consider $S=10$ and $S=1$ sectors. For $S=1$, there is effectively no divide component in the DC algorithm, thus making it equivalent to the traditional approach to the IM problem considered in the literature~\cite{erkol2019systematic}. In Fig.~\ref{fig:4} we compare directly the metrics of performance of Eq.~(\ref{eq:norm_auc}) obtained with $S=10$ and $S=1$ over the entire corpus of real networks. The ratios for $S=1$ are indicated $R_\textrm{a}$, $R_\textrm{c}$ and $R_\textrm{e}$, for adaptive degree, collective influence and eigenvector  centrality, respectively; for $S=10$, the ratios are instead indicated as $R_\textrm{Pa}$, $R_\textrm{Pc}$ and $R_\textrm{Pe}$.  The scatter plots show that if we follow the divide and conquer strategy higher scores are obtained than if influencers are picked from the network as a whole. This holds true regardless of the centrality metric used to proxy the influence of the individual nodes. 

Finally, we study how the performance of the DC algorithm depends on the type of method implemented to divide the network into sectors.  We find that $R_\textrm{Pa} \geq R_\textrm{Ca}$ for $44$ out of $52$ real networks, meaning that graph partitioning is better suited than community structure to define sectors of influence in a real network (Fig.~\ref{fig:5}a). The same result holds for the comparison $R_\textrm{Pa}$ vs. $R_\textrm{Ea}$ (Fig.~\ref{fig:5}b). Graph embedding and community structure yield instead similar performance (Fig.~\ref{fig:5}c).

\subsection{Synthetic networks}

We generate LFR networks with $N = 1,000$ nodes~\cite{lancichinetti2008benchmark}.
We vary the mixing parameter $\mu$ from $0.05$ to $0.40$ to control for the strength of the planted community structure and the degree exponent $\tau_1$ from $1.7$ to $4.0$ to tune the heterogeneity of the degree distribution.  
We set the community size power-law exponent $\tau_2 = 1.0$, the average degree $\langle k \rangle = 10$, and the maximum degree $k_{max} = 70$. 

For each network, we identify the best set of seed nodes using three different strategies. Two of these strategies do not involve the division of the network in any sectors; we simply identify the top spreaders via adaptive degree centrality and collective influence with $\ell = 2$ in the entire network. The third strategy takes advantage of the DC algorithm with $S=10$ sectors defined using graph partitioning; top influencers are identified based on adaptive degree centrality on the various sectors. In Figure~\ref{fig:6}, we display the ratios $R_\textrm{Pa}/R_\textrm{a}$ and $R_\textrm{Pa}/R_\textrm{c}$ as functions of the mixing parameter $\mu$ of the model. Results are obtained by averaging the ratios over $50$ realizations of the network model and of the procedure for the identification of the spreaders. We report results for different values of the degree exponent $\tau_1$. As in the case of real networks, dividing the network into sectors allows us to obtain better solutions to the IM problem than those obtained without any division. The gain in performance increases as the degree heterogeneity of the nodes and the strength of the modular structure of the network increase.

\section{Discussion} 

We have proposed a two-step strategy to search for 
effective influencers in networks. By dividing the graph into sectors and finding influencers independently in each sector, via widely adopted centrality scores, we showed that it is possible to increase the relative outbreak size with respect to algorithms sorting nodes based on their centrality in the whole network. The improvement is the larger, the more modular the graph is and the more heterogeneous its degree distribution is. The gain produced by our distributed approach does not come at the expenses of the time complexity of the procedure, as the division of the network into (a constant number of) sectors can be done in linear time, so the total complexity is dominated by the calculation of the centrality scores.
Our numerical experiments show that graph partitioning techniques are highly effective at identifying the sectors.

\newpage

\section*{Acknowledgement}
We thank \c{S}irag Erkol for useful advice in the development of the spreading maximization routine.
This project was partially supported by the Army Research Office under contract number W911NF-21-1-0194, by the Air Force Office of Scientific Research under award numbers FA9550-19-1-0391 and FA9550-21-1-0446, and by the National Science Foundation under award number 1927418. The
funders had no role in study design, data collection and
analysis, decision to publish, or any opinions, findings,
and conclusions or recommendations expressed in the
manuscript.

\section*{Appendix}

Table \ref{table1} summarizes the information of the 52 networks considered in the corpus. We report the name of the network, its type, the number of nodes and edges in the giant component, the critical percolation threshold, references to studies where the network is presented and analyzed, and the {\tt url} for the data.

\begin{table*}[!htb]
\begin{center}
\centering
\begin{tabular}{|l|c|r|r|c|c|c|c|}\hline
\bf{Network} & \bf{Type} & \bf{N} & \bf{E} & $\bf{p^*}$ & \bf{Ref.} & \bf{url} \\\hline
US Air Trasportation & transportation & 500 & 2980 & 0.026 & \cite{colizza2007reaction} & \href{https://sites.google.com/site/cxnets/usairtransportationnetwork}{url} \\\hline
URV email & social & 1133 & 5451 & 0.056 & \cite{guimera2003self} & \href{http://deim.urv.cat/~alexandre.arenas/data/welcome.htm}{url} \\\hline
Political blogs & information & 1222 & 16714 & 0.015 & \cite{adamic2005political} & \href{http://www-personal.umich.edu/~mejn/netdata/}{url} \\\hline
Air traffic & transportation & 1226 & 2408 & 0.163 & \cite{konect} & \href{http://konect.uni-koblenz.de/networks/maayan-faa}{url} \\\hline
Petster, hamster & social & 1788 & 12476 & 0.025 & \cite{konect} & \href{http://konect.uni-koblenz.de/networks/petster-friendships-hamster}{url} \\\hline
UC Irvine & social & 1893 & 13835 & 0.023 & \cite{opsahl2009clustering} & \href{http://konect.uni-koblenz.de/networks/opsahl-ucsocial}{url} \\\hline
Yeast, protein & biological & 2224 & 6609 & 0.071 & \cite{bu2003topological} & \href{http://vlado.fmf.uni-lj.si/pub/networks/data/bio/Yeast/Yeast.htm}{url} \\\hline
Adolescent health & social & 2539 & 10455 & 0.117 & \cite{moody2001peer,icon} & \href{http://konect.uni-koblenz.de/networks/moreno_health}{url} \\\hline
USFCA & social & 2672 & 65244 & 0.011 & \cite{traud2012social,traud2011fs,nr} & \href{http://networkrepository.com/socfb-USFCA72.php}{url} \\\hline
Japanese & information & 2698 & 7995 & 0.030 & \cite{milo2004superfamilies} & \href{http://wws.weizmann.ac.il/mcb/UriAlon/index.php?q=download/collection-complex-networks}{url} \\\hline
Open flights & transportation & 2905 & 15645 & 0.020 & \cite{opsahl2010node,konect} & \href{http://konect.uni-koblenz.de/networks/opsahl-openflights}{url} \\\hline
Pepperdine & social & 3440 & 152003 & 0.007 & \cite{traud2012social,traud2011fs,nr} & \href{http://networkrepository.com/socfb-Pepperdine86.php}{url} \\\hline
Wesleyan & social & 3591 & 138034 & 0.009 & \cite{traud2012social,traud2011fs,nr} & \href{http://networkrepository.com/socfb-Wesleyan43.php}{url} \\\hline
Mich & social & 3745 & 81901 & 0.011 & \cite{traud2012social,traud2011fs,nr} & \href{http://networkrepository.com/socfb-Mich67.php}{url} \\\hline
Bitcoin Alpha & social & 3775 & 14120 & 0.027 & \cite{kumar2016edge,kumar2018rev2,snapnets} & \href{http://snap.stanford.edu/data/soc-sign-bitcoin-alpha.html}{url} \\\hline
Bucknell & social & 3824 & 158863 & 0.008 & \cite{traud2012social,traud2011fs,nr} & \href{http://networkrepository.com/socfb-Bucknell39.php}{url} \\\hline
Howard & social & 4047 & 204850 & 0.006 & \cite{traud2012social,traud2011fs,nr} & \href{http://networkrepository.com/socfb-Howard90.php}{url} \\\hline
GR-QC, 1993-2003 & social & 4158 & 13422 & 0.091 & \cite{leskovec2007graph,snapnets} & \href{http://snap.stanford.edu/data/ca-GrQc.html}{url} \\\hline
Tennis & social & 4338 & 81865 & 0.007 & \cite{radicchi2011best} & - \\\hline
US Power grid & technological & 4941 & 6594 & 0.437 & \cite{watts1998collective} & \href{http://www-personal.umich.edu/~mejn/netdata/}{url} \\\hline
HT09 & social & 5352 & 18481 & 0.025 & \cite{isella2011s} & \href{http://www.sociopatterns.org/datasets/hypertext-2009-dynamic-contact-network/}{url} \\\hline
Hep-Th, 1995-1999 & social & 5835 & 13815 & 0.108 & \cite{newman2001structure} & \href{http://www-personal.umich.edu/~mejn/netdata/}{url} \\\hline
Bitcoin OTC & social & 5875 & 21489 & 0.023 & \cite{kumar2016edge,kumar2018rev2,snapnets} & \href{http://snap.stanford.edu/data/soc-sign-bitcoin-otc.html}{url} \\\hline
Reactome & biological & 5973 & 145778 & 0.011 & \cite{joshi2005reactome,konect} & \href{http://konect.uni-koblenz.de/networks/reactome}{url} \\\hline
Jung & technological & 6120 & 50290 & 0.009 & \cite{vsubelj2012software,konect} & \href{http://konect.uni-koblenz.de/networks/subelj_jung-j}{url} \\\hline
Gnutella, Aug. 8, 2002 & technological & 6299 & 20776 & 0.046 & \cite{ripeanu2002mapping,leskovec2007graph,snapnets} & \href{http://snap.stanford.edu/data/p2p-Gnutella08.html}{url} \\\hline
JDK & technological & 6434 & 53658 & 0.009 & \cite{konect} & \href{http://konect.uni-koblenz.de/networks/subelj_jdk}{url} \\\hline
UChicago & social & 6561 & 208088 & 0.008 & \cite{traud2012social,traud2011fs,nr} & \href{http://networkrepository.com/socfb-UChicago30.php}{url} \\\hline
UC & social & 6810 & 155320 & 0.010 & \cite{traud2012social,traud2011fs,nr} & \href{http://networkrepository.com/socfb-UC64.php}{url} \\\hline
Wikipedia elections & social & 7066 & 100736 & 0.008 & \cite{leskovec2010signed,leskovec2010predicting,snapnets} & \href{http://snap.stanford.edu/data/wiki-Vote.html}{url} \\\hline
English & information & 7377 & 44205 & 0.011 & \cite{milo2004superfamilies} & \href{http://wws.weizmann.ac.il/mcb/UriAlon/index.php?q=download/collection-complex-networks}{url} \\\hline
Gnutella, Aug. 9, 2002 & technological & 8104 & 26008 & 0.045 & \cite{ripeanu2002mapping,leskovec2007graph,snapnets} & \href{http://snap.stanford.edu/data/p2p-Gnutella09.html}{url} \\\hline
French & information & 8308 & 23832 & 0.022 & \cite{milo2004superfamilies} & \href{http://wws.weizmann.ac.il/mcb/UriAlon/index.php?q=download/collection-complex-networks}{url} \\\hline
Hep-Th, 1993-2003 & social & 8638 & 24806 & 0.072 & \cite{leskovec2007graph,snapnets} & \href{http://snap.stanford.edu/data/ca-HepTh.html}{url} \\\hline
Gnutella, Aug. 6, 2002 & technological & 8717 & 31525 & 0.065 & \cite{ripeanu2002mapping,leskovec2007graph,snapnets} & \href{http://snap.stanford.edu/data/p2p-Gnutella06.html}{url} \\\hline
Gnutella, Aug. 5, 2002 & technological & 8842 & 31837 & 0.056 & \cite{ripeanu2002mapping,leskovec2007graph,snapnets} & \href{http://snap.stanford.edu/data/p2p-Gnutella05.html}{url} \\\hline
PGP & social & 10680 & 24316 & 0.064 & \cite{boguna2004models} & \href{http://deim.urv.cat/~alexandre.arenas/data/welcome.htm}{url} \\\hline
Gnutella, Aug. 4, 2002 & technological & 10876 & 39994 & 0.076 & \cite{ripeanu2002mapping,leskovec2007graph,snapnets} & \href{http://snap.stanford.edu/data/p2p-Gnutella04.html}{url} \\\hline
Hep-Ph, 1993-2003 & social & 11204 & 117619 & 0.005 & \cite{leskovec2007graph,snapnets} & \href{http://snap.stanford.edu/data/ca-HepPh.html}{url} \\\hline
Spanish 1 & information & 11558 & 43050 & 0.012 & \cite{milo2004superfamilies} & \href{http://wws.weizmann.ac.il/mcb/UriAlon/index.php?q=download/collection-complex-networks}{url} \\\hline
DBLP, citations & information & 12495 & 49563 & 0.032 & \cite{ley2002dblp,konect} & \href{http://konect.uni-koblenz.de/networks/dblp-cite}{url} \\\hline
Spanish 2 & information & 12643 & 55019 & 0.012 & \cite{konect} & \href{http://konect.uni-koblenz.de/networks/lasagne-spanishbook}{url} \\\hline
Cond-Mat, 1995-1999 & social & 13861 & 44619 & 0.064 & \cite{newman2001structure,snapnets} & \href{http://www-personal.umich.edu/~mejn/netdata/}{url} \\\hline
Astrophysics & social & 14845 & 119652 & 0.018 & \cite{newman2001structure} & \href{http://www-personal.umich.edu/~mejn/netdata/}{url} \\\hline
AstroPhys, 1993-2003 & social & 17903 & 196972 & 0.013 & \cite{leskovec2007graph,snapnets} & \href{http://snap.stanford.edu/data/ca-AstroPh.html}{url} \\\hline
Cond-Mat, 1993-2003 & social & 21363 & 91286 & 0.037 & \cite{leskovec2007graph,snapnets} & \href{http://snap.stanford.edu/data/ca-CondMat.html}{url} \\\hline
Gnutella, Aug. 25, 2002 & technological & 22663 & 54693 & 0.115 & \cite{ripeanu2002mapping,leskovec2007graph,snapnets} & \href{http://snap.stanford.edu/data/p2p-Gnutella25.html}{url} \\\hline
Internet & technological & 22963 & 48436 & 0.019 & None & \href{http://www-personal.umich.edu/~mejn/netdata/}{url} \\\hline
Thesaurus & information & 23132 & 297094 & 0.011 & \cite{kiss1973associative,konect} & \href{http://konect.uni-koblenz.de/networks/eat}{url} \\\hline
Cora & information & 23166 & 89157 & 0.045 & \cite{vsubelj2013model,konect} & \href{http://konect.uni-koblenz.de/networks/subelj_cora}{url} \\\hline
AS Caida & technological & 26475 & 53381 & 0.021 & \cite{leskovec2005graphs,snapnets} & \href{http://snap.stanford.edu/data/as-caida.html}{url} \\\hline
Gnutella, Aug. 24, 2002 & technological & 26498 & 65359 & 0.106 & \cite{ripeanu2002mapping,leskovec2007graph,snapnets} & \href{http://snap.stanford.edu/data/p2p-Gnutella24.html}{url} \\\hline
\end{tabular}
\end{center}
\caption{List of the real networks analyzed in the study. From left to right we report the name of the network, its type, the number of nodes in the giant component, the number of edges in the giant component, the percolation threshold, references to studies where the network is presented and analyzed, and the {\tt url} where the network can be found.}
\label{table1}
\end{table*}

\clearpage

\bibliography{bibliography.bib}

\end{document}